\documentclass[prb,twocolumn,groupedaddress,superscriptaddress]{revtex4}
\usepackage{graphicx}
\usepackage{amssymb}
\usepackage{latexsym}
\usepackage[usenames]{color}
\usepackage{float}

\input{tcilatex}

\begin{document}

\title{New Perspectives for Cuprate Research: (Ca$_{x}$La$_{1-x}$)(Ba$%
_{1.75-x}$La$_{0.25+x}$)Cu$_{3}$O$_{y}$ Single Crystals}
\author{Gil Drachuck}
\affiliation{Department of Physics, Technion - Israel Institute of Technology, Haifa
32000, Israel}
\author{Meni Shay}
\affiliation{Department of Physics and Optical Engineering, Ort Braude College, P.O. Box
78, 21982 Karmiel, Israel}
\author{Galina Bazalitsky}
\affiliation{Department of Physics, Technion - Israel Institute of Technology, Haifa
32000, Israel}
\author{Rinat Ofer}
\affiliation{Department of Physics, Technion - Israel Institute of Technology, Haifa
32000, Israel}
\author{Zaher Salman}
\author{Alex Amato}
\affiliation{Laboratory for Muon Spectroscopy, Paul Scherrer Institute, CH-5232 Villigen
PSI, Switzerland}
\author{Christof Niedermayer}
\affiliation{Laboratory for Neutron Scattering, Paul Scherrer Institute, CH-5232 Villigen
PSI, Switzerland}
\author{Dirk Wulferding}
\author{Peter Lemmens}
\affiliation{Institute for Condensed Matter Physics, TU Braunschweig, D-38106
Braunschweig, Germany}
\author{Amit Keren}
\affiliation{Department of Physics, Technion - Israel Institute of Technology, Haifa
32000, Israel}

\begin{abstract}
We report the successful growth of a large (Ca$_{x}$La$_{1-x}$)(Ba$_{1.75-x}$La%
$_{0.25+x}$)Cu$_{3}$O$_{y}$ (CLBLCO) single crystal. In this material, $x$
controls the maximum of $T_{c}$ ($T_{c}^{\max }$), with minimal structural
changes. Therefore, it allows a search for correlations between material
properties and $T_{c}^{\max }$. We demonstrate that the crystals are good
enough for neutron scattering and cleave well enough for Raman scattering.
These results open new possibilities for cuprate research.
\end{abstract}

\maketitle

Although high quality single crystals of cuprate superconductors have been
available for quite some time, comparing the properties of different
crystals often raised more questions than answers. The main problem is that
many parameters differ between various cuprates simultaneously, and it is
difficult to understand which material property is responsible for which
physical property. A proper understanding of cuprate superconductivity can
emerge only when it is possible to vary the material parameters one at a
time preferably in single crystals. Here we describe the growth and
characterization of large single crystals of a (Ca$_{x}$La$_{1-x}$)(Ba$%
_{1.75-x}$La$_{0.25+x}$)Cu$_{3}$O$_{y}$ (CLBLCO) superconductor, in which it
was previously demonstrated that the variable $x$\ changes only the Cu-O-Cu
buckling angle and bond distance, and hence the super-exchange \cite%
{OferPRB08}. We also demonstrate that experiments such as neutron
scattering, Raman scattering, and more can be performed on these crystals.

\begin{figure}[h]
\centering \includegraphics[width=7.2cm]{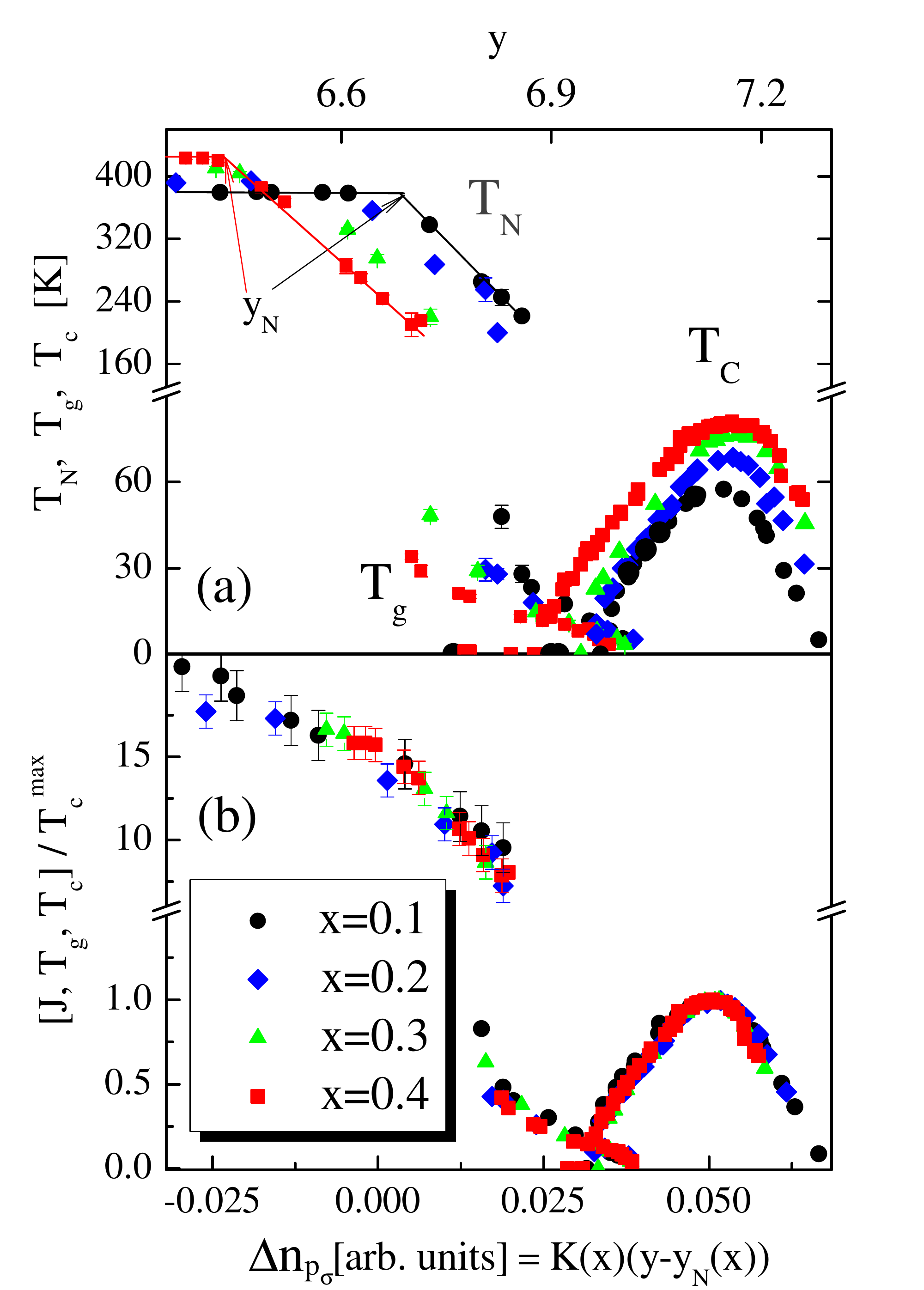}
\caption{ (a) The phase diagram of CLBLCO showing the N\'{e}el ($T_{N}$),
glass ($T_{g}$) and superconducting ($T_{c}$) temperatures over the full
doping range for the four families. $y_{N}$ indicates the oxygen level where
$T_{N}$ start to drop. (b) The unified phase diagram of CLBLCO. The critical
temperatures, and $J$ extracted from $T_{N}$, are divided by $T_{c}^{max}$
and plotted as a function of doping variation in the oxygen orbital $\Delta
n_{p_{\protect\sigma }}$.}
\label{fig:PhaseDiagram}
\end{figure}

The phase diagram of CLBLCO for various values of $x$ and $y$ is presented
in Fig.~\ref{fig:PhaseDiagram}(a) \cite{AmitPRB10}. When varying $x$, the
amount of Lanthanum in the chemical formula remains constant, and therefore $%
x$ stands for the Calcium-to-Barium ratio. The parameter $y$ controls the
oxygen level and moves the system between the different phases. At around $%
y=7.15$ each family has its maximum $T_{c}$ [$T_{c}^{max}$]. Changing $x$
from 0.1 to 0.4 varies $T_{c}^{max}$ from 58~K to 80~K~\cite{Yaki99};
roughly a 30\% increase. This $T_{c}^{max}$ variation is achieved with no
apparent structural changes.

All CLBLCO compounds have YBCO-like structure with two CuO$_{2}$ planes and
two disordered \textquotedblleft chain" layers per unit cell. The symmetry
is tetragonal for all values of $x$ and $y$ \cite{KellerPhysicaC89}. The
level of disorder is also similar for all families \cite{KerenNJP09}.
Therefore, \emph{a priori}, there is no reason for $T_{c}^{max}$ to depend
on $x$. However, increasing $x$ increases the amount of the Ca$^{2+}$ in the
Y site of YBCO at the expense of La$^{3+}$. This charge transfer is
equivalent to reducing the positive charge on the (YBCO) Y site and
increasing it on the (YBCO) Ba site, and could alter the Cu-O-Cu buckling
angle. Indeed, it was found by high resolution neutron diffraction that as $x
$ increases the Cu-O-Cu bond becomes shorter and straighter \cite{OferPRB08}%
, while other structural properties remain intact \cite{KerenNJP09}. The
bond length and buckling are the major factors that control the orbital
overlap and, in turn, the hopping parameter $t$ or the super-exchange $J$.
Thus, the newly grown CLBLCO crystal can shed new light on the impact of $t$
or $J$ on properties such as: $T_{c}^{max}$, the structure of the Fermi
surface, the magnetic resonance, the size of the gap/pseudogap, and more.

For example, early muon spin rotation ($\mu $SR) measurements using
compressed CLBLCO powder samples in the anti-ferromagnetic phase, combined
with theoretical calculations, revealed that $J$ is $x$-dependent, as
expected, and that $T_{c}^{\max }(x)\propto J(x)$ \cite{OferPRB06}.
Similarly, NMR established that simple valence counting does not represent
the true doping of the CuO$_{2}$ planes and that a doping efficiency
parameter $K(x)$ should be introduced; the number of holes in the oxygen 2p$%
_{\sigma }$ orbital is given by $\Delta n_{2p}=K(x)(y-y_{N})$ where $y_{N}$
is defined in Fig.~\ref{fig:PhaseDiagram}(a) \cite{AmitPRB10}.
Interestingly, when plotting $J(x)$, $T_{g}$, and $T_{c}$ normalized by $%
T_{c}^{\max }(x)$ as a function of $\Delta n_{2p}$ a unified phase diagram
is generated, as depicted in Fig.~\ref{fig:PhaseDiagram}(b) \cite%
{OferPRB06,AmitPRB10}.

\begin{figure}[h]
\centering \includegraphics[width=6cm]{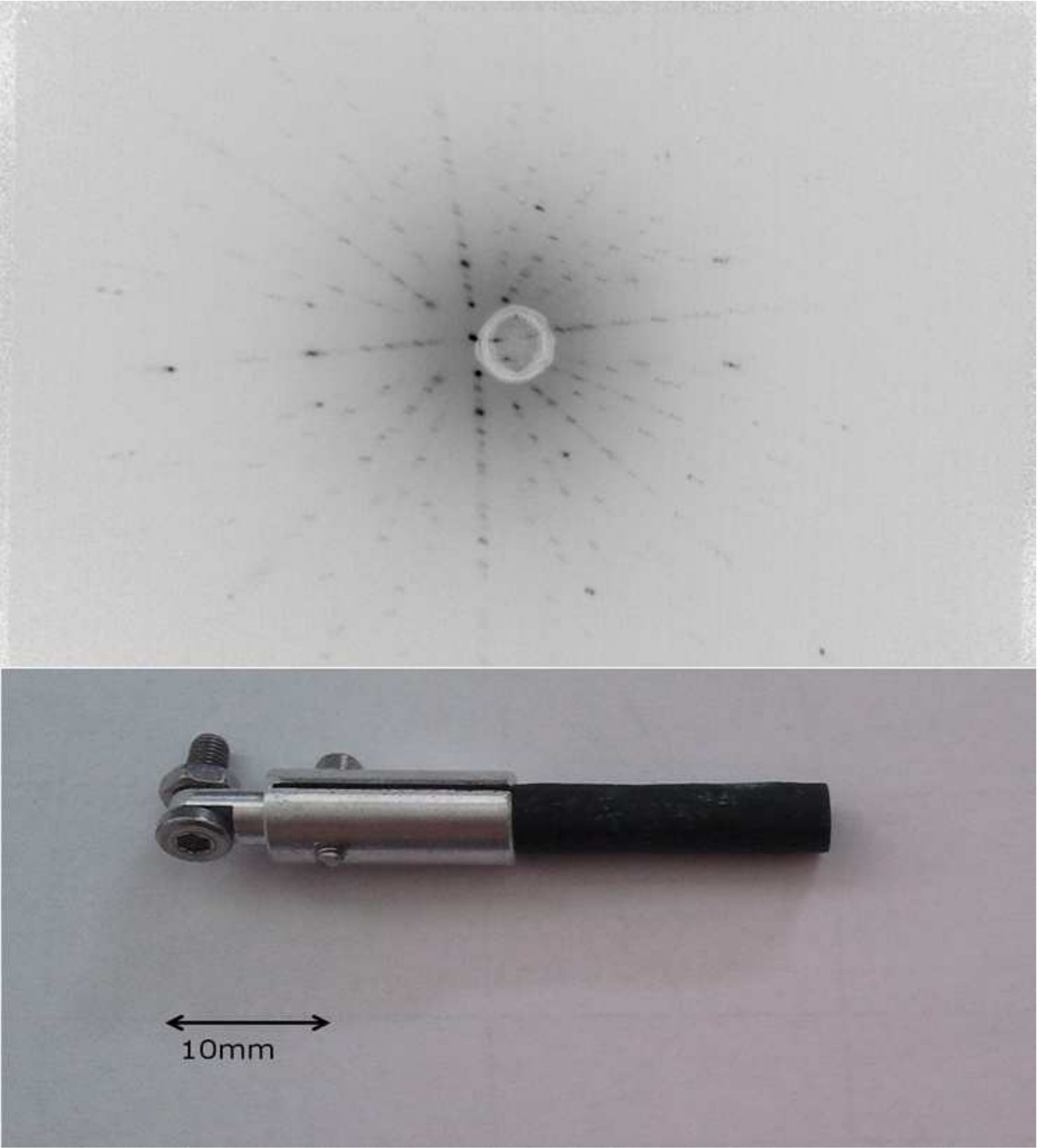}
\caption{Top: Laue Diffraction pattern of the (001) crystal plane taken from
the CLBLCO crystal. Bottom: Photograph of the grown CLBLCO Crystal.}
\label{fig:Crystal}
\end{figure}

However, $\mu $SR is not a direct probe of $J,$ and more profound work is
needed to clarify the relation between the superconducting and magnetic
energy scales. Inelastic neutron or Raman scattering are more appropriate
for direct $J$ determination. Equally, NMR does not measure the Fermi
surface size directly and a proper ARPES measurement is needed. These
advanced measurements require large single crystals with specific
orientation or high quality surfaces. Here we report the production of such
crystals by the Traveling Solvent Floating Zone method. Details of the
growth and characterization are given in the Appendix material. An optical
image and a Laue diffraction pattern of one of the crystals are shown in
Fig.~\ref{fig:Crystal}. We also demonstrate the applicability of several
experimental techniques to these crystals. It should be pointed out that
small crystals were grown before but only their chemical properties where
analyzed \cite{Noji99}.

\begin{figure}[h]
\centering \includegraphics[width=7.2cm]{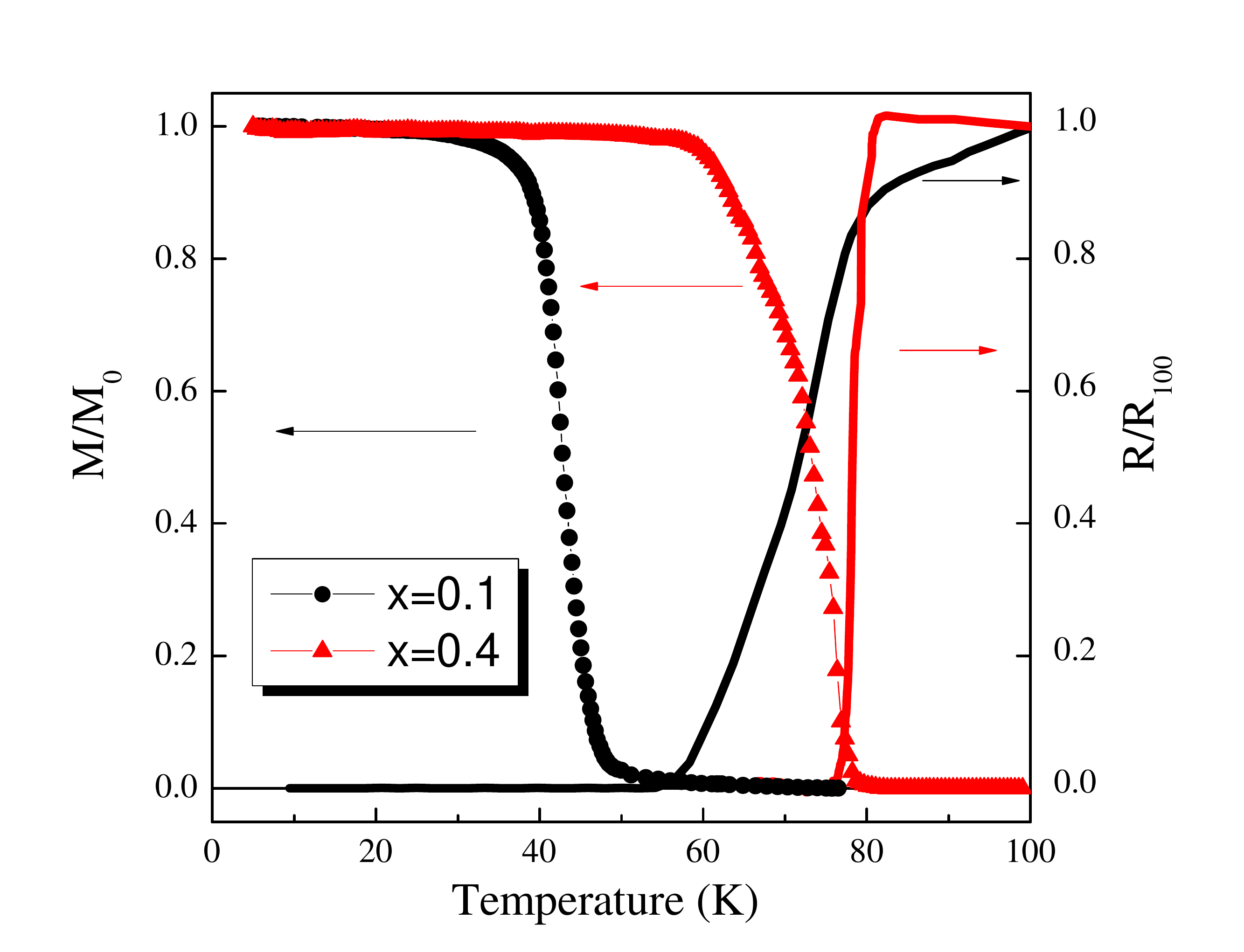}
\caption{ Normalized resistivity and susceptibility as a function of $T$
obtained from an optimally doped crystal with $x=0.1$ and $x=0.4$.}
\label{fig:Bulk}
\end{figure}

The bulk superconducting properties of this crystal were tested by transport
and magnetization measurements. The $x=0.1$ and $0.4$ crystals are cut into
a rectangular shape and post-annealed in oxygen to achieve optimal doping.
The $x=0.1$ is annealed at a temperature of 300~C$^{\circ }$ and ambient
pressure for $150$~hr. The $x=0.4$ is annealed at a temperature of $400$~C$%
^{\circ }$ and a pressure of $65$~bar for $90$~hr. Figure~\ref{fig:Bulk}
shows normalized resistivity and normalized susceptibility. We found that $%
T_{c}=58$ and $78$~K for $x=0.1$ and $0.4$ respectively, which is in perfect
agreement with powder sample measurements. $T_{c}$ is defined as the
temperature where the resistivity vanishes upon cooling or the
susceptibility vanishes upon warming. The transition into the
superconducting state of the $x=0.1$ crystal is wide. It spans $10$~K in the
susceptibility measurement and $25$ K in the resistivity measurement. This
is probably due to very small traces of high $x$ values. The transition of
the $x=0.4$ crystal is sharp in both susceptibility and resistivity
measurements. Nevertheless, there are clear variations in $T_{c}$ between
optimally doped crystal with different $x$ values. The superconducting
volume fraction extracted from the magnetization measurement is
approximately 100\%.

\begin{figure}[h]
\centering \includegraphics[width=8cm]{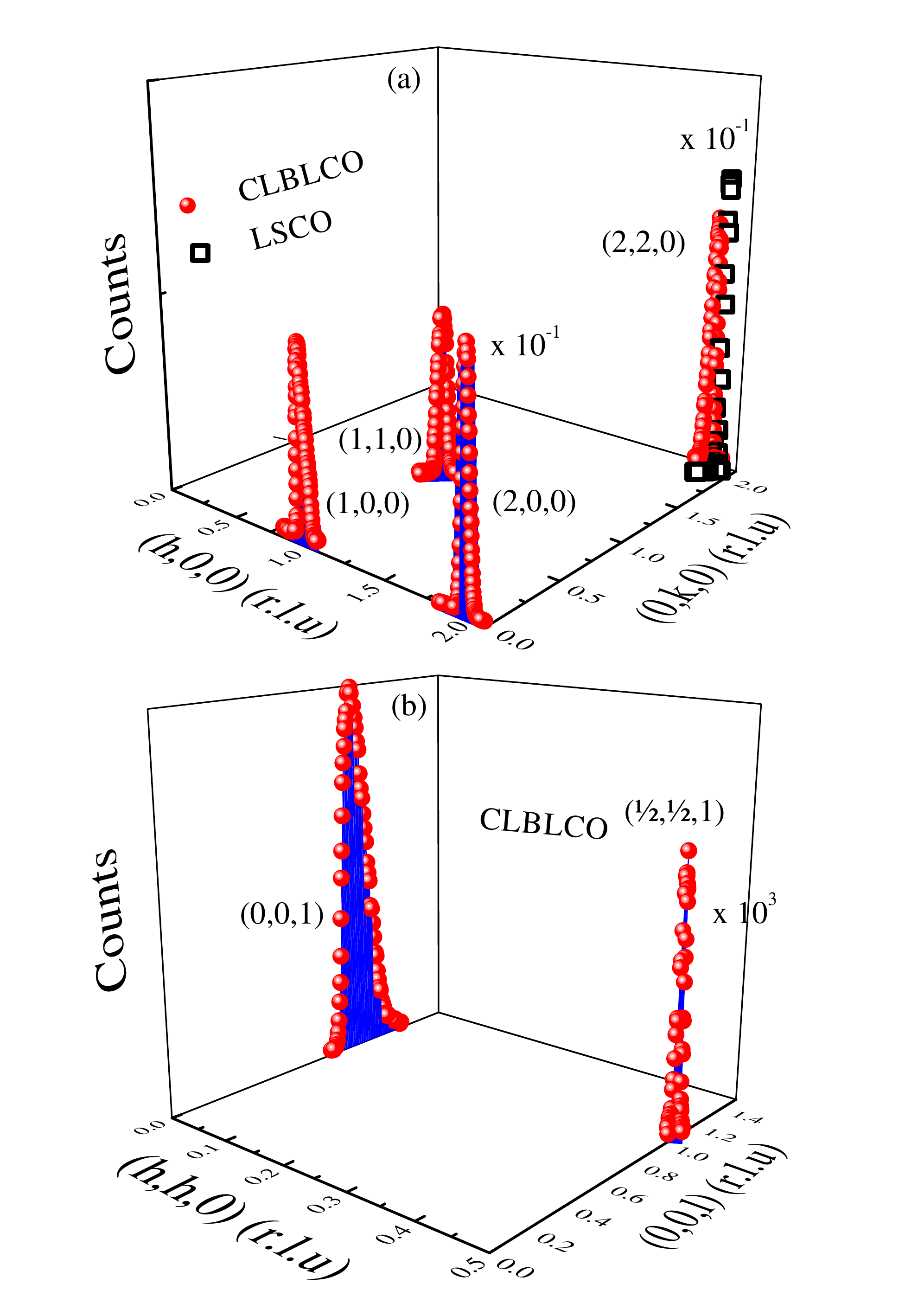}
\caption{(a) The structural Bragg peaks (red circles) of the CLBLCO $x=0.1$
single crystal obtained by neutron scattering. The (200) peak intensity is
divided by 10 for clarity. The empty black squares described the (020) peak
measure on a LSCO crystal as a reference. (b) The magnetic Bragg peak ($%
\frac{1}{2}\frac{1}{2}$1) multiplied by $10^{3}$ for clarity, and the
structural peak (001). }
\label{fig:NeutronPeaks}
\end{figure}

The most demanding requirements from single crystals are those of the
neutron scattering experimental technique. For this technique the crystals
must be of the order of 10~gr. To check the applicability of neutron
scattering to our crystal we use the ORION beam-line at PSI, Switzerland.
This beam-line provides neutrons with a wavelength of $\lambda =2.22$\ $%
\mathring{A}$. The crystal is kept at room temperature. Using an (h,k,0)
scan and the YBCO structure parameters taken from ISCD's \textquotedblleft
Lazy Pulverix" website, we detect the (100), (200), (110), (220) nuclear
Bragg peaks as presented in Fig.~\ref{fig:NeutronPeaks}(a). For comparing
crystal qualities we also present the strongest LSCO nuclear Bragg peak,
namely, the (220) from a sample of equal size. The intensity of the
strongest LSCO peak, (2,2,0), and the strongest CLBLCO peak, (2,0,0), is
similar.

In Fig~\ref{fig:NeutronPeaks}(b) we show neutron scattering data acquired on
the MORPHEUS beam-line. This beam line has a higher flux and a wavelength of
$\lambda =5.02$ $\mathring{A}$ which enables us access to the first allowed
magnetic Bragg peak of CLBLCO. Indeed, using an (h,h,l) scan we are able to
detect the nuclear $(001)$ and the magnetic $(\frac{1}{2}\frac{1}{2}1)$
Bragg peaks of our CLBLCO crystal as well. The magnetic peak intensity in
the figure is multiplied by $10^{3}$ for clarity.
\begin{figure}[t]
\centering \includegraphics[height=10cm]{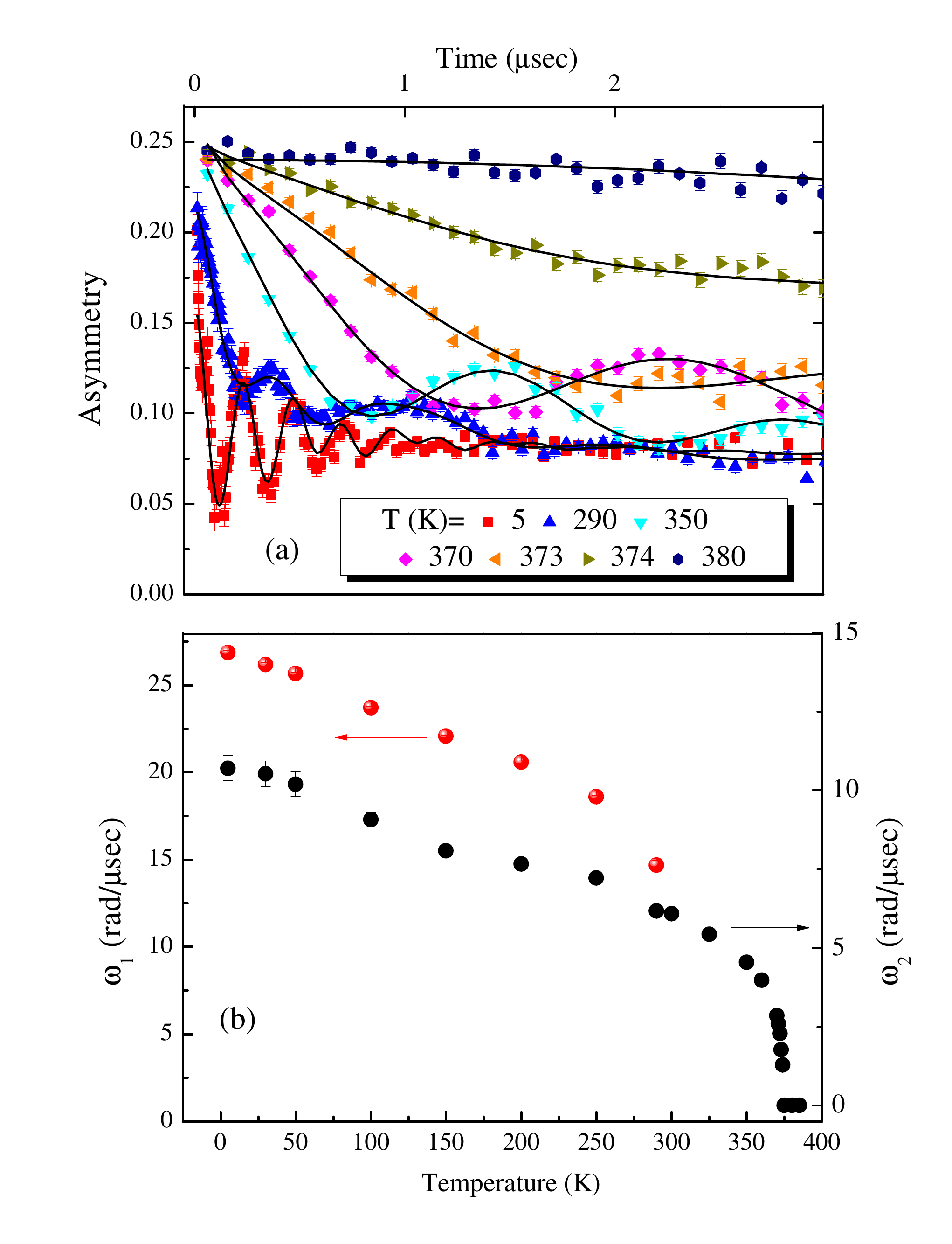}
\caption{ (a): $\protect\mu $SR Asymmetry with a fit of Eq.~\protect\ref%
{Asymmetry} to the data at various temperatures.(b): muon spin oscillation
at high frequency $\protect\omega _{1}$ (red) and low frequency $\protect%
\omega _{2}$ (black) as a function of temperature from the CLBLCO $x=0.1$
crystal. }
\label{fig:MuSRFrequency}
\end{figure}

We also check the major magnetic properties of our crystal by performing a
ZF-$\mu $SR experiment, at PSI, Switzerland. Figure~\ref{fig:MuSRFrequency}%
(a) shows the $\mu $SR asymmetry data of an $x=0.1$ crystal by symbols, and
the fit of the function
\begin{equation}
{Asy(}t)=\sum\limits_{i=0}^{2}{{A_{i}}}\exp (-{\lambda _{i}}t)\cos ({\omega
_{i}}t+\phi )+{B_{g}}  \label{Asymmetry}
\end{equation}%
to the data by solid lines. Here $\omega _{0}$ is fixed at $0$, and $\omega
_{1}$ and $\omega _{2}$ stand for high and low frequencies, respectively.
The two frequencies extracted from the fit are presented in Fig~\ref%
{fig:MuSRFrequency}(b). They result from two different muon sites, common to
most cuprates. Above 300~K the two frequencies merge into one. The N\'{e}el
temperature is determined from the $\omega \rightarrow 0$ limit giving $%
T_{N}=375~$K, which is the same as the $x=0.1$ powder sample.

\begin{figure}[h]
\centering \includegraphics[width=8cm]{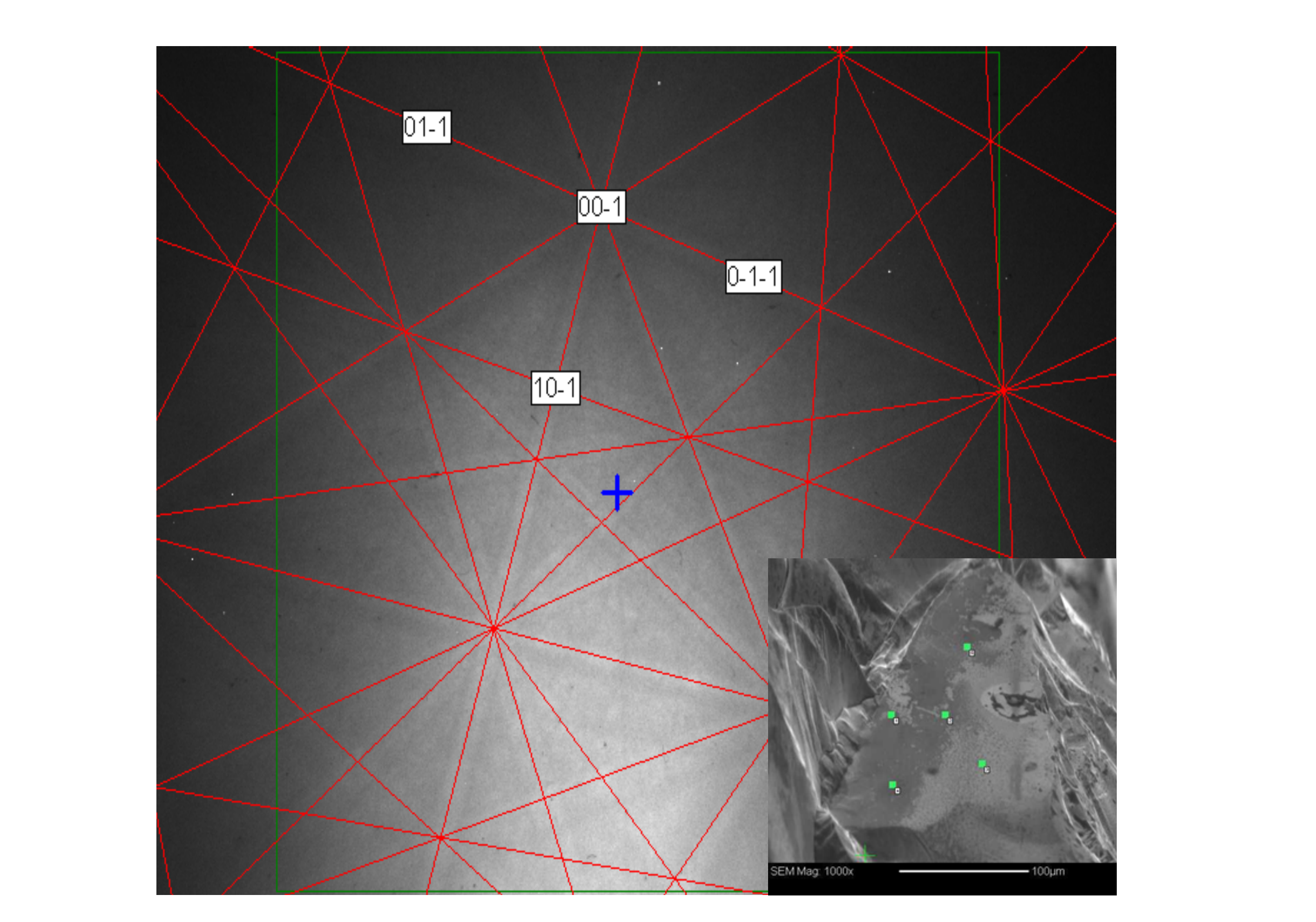}
\caption{ Main panel: The EBSD diffraction pattern consisting of
intersecting Kikuchi lines, where the red lines depict the most probable fit
to the YBCO structure. Inset: SEM micrograph of the CLBLCO crystal showing a
large facet. The Green squares identifies the spots at which EBSD patterns
were acquired. }
\label{fig:SEM}
\end{figure}

For a variety of experimental techniques having single crystals is
insufficient; these crystals must also cleave with high quality exposed
surfaces. To test this requirement, we cleave one of our samples and examine
the surface by a scanning electron microscope (SEM) equipped with an
electron backscatter diffraction module (EBSD). A SEM micrograph taken at an
acceleration voltage of 25~kV is shown in the inset of Fig.~\ref{fig:SEM}. A
100x200 $\mu m^{2}$ crystal facet is clearly visible. The EBSD diffraction
patterns from one point on the surface is shown in Fig.~\ref{fig:SEM}. Clear
Kikuchi lines can be seen in the diffraction \cite{kikichi}. A fit to the
expected Kikuchi lines from the YBCO structure is also presented. The
patterns reveal that the surface of the facet is perpendicular to the
crystal $\mathbf{c}$-axis. We repeat the EBSD measurements at all the
(green) points in the inset, and find the same Kikuchi pattern. This
indicates that all points share the same crystal orientation and that the
surface is perpendicular to the crystallographic $\mathbf{c}$-axis. This
finding is also confirmed by the Raman scattering experiment described below.

\begin{figure}[tbp]
\centering \includegraphics[height=6.8cm]{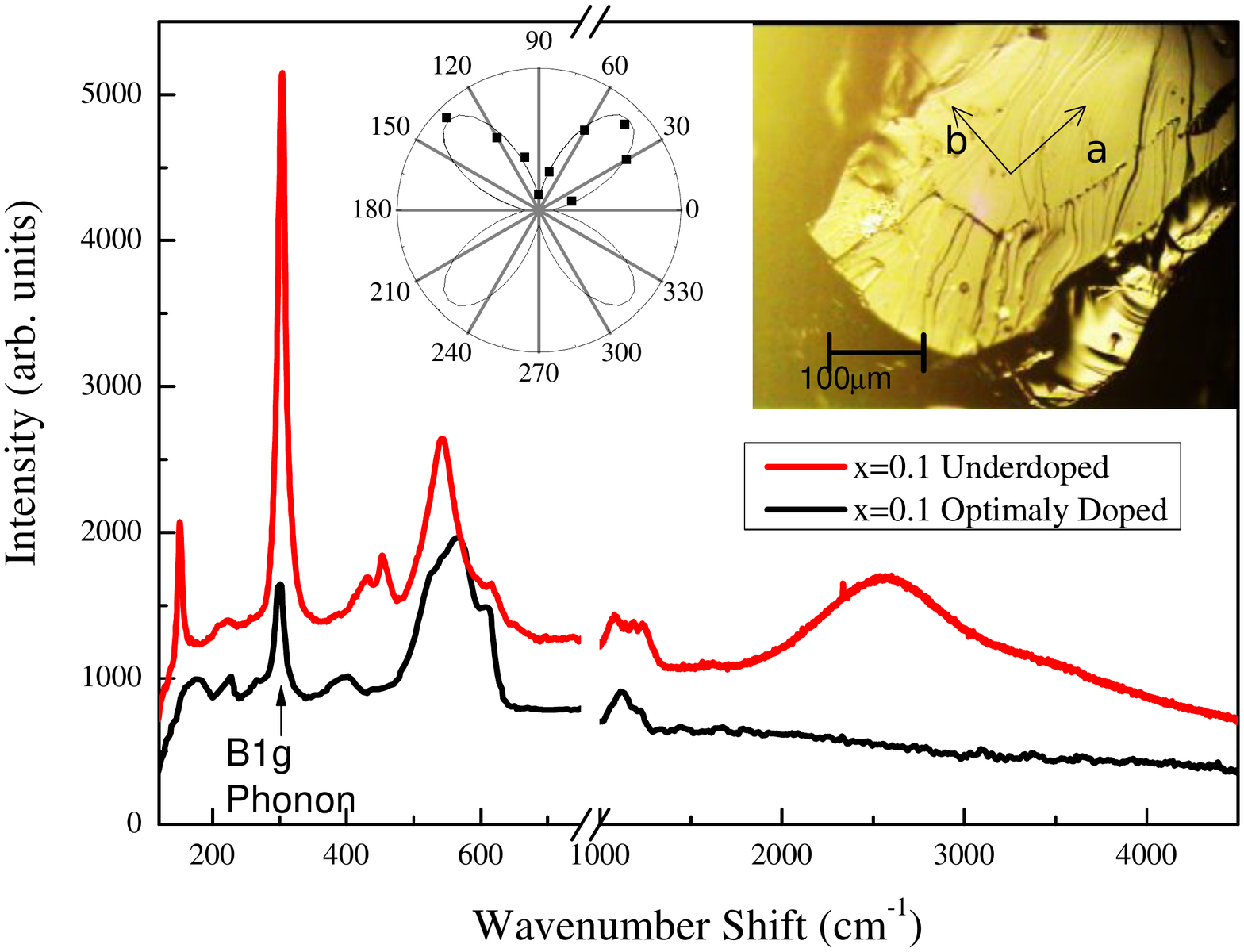}
\caption{Main panel: Unpolarized Raman spectra of the underdoped crystal
(red) and the optimally doped crystal (black); note the change of scale at
the axis breaker. The optical phonons are in the range $100-600$ cm$^{-1}$.
The broad two-magnon peak is at $2500$~cm$^{-1}$. This peak exists only in
the underdoped crystal, as expected. Left inset: The intensity of the $%
\mathrm{{B_{1g}}}$ phonon as a function of the angle between the incident
polarization and the crystal a-axis using polarizer and analyzer at 90$%
^{\circ }$ to the polarizer. This shows the 4-fold symmetry of the crystal
and determines its orientation. Right inset: The cleaved crystal facet and
the crystal axis as determined using the $\mathrm{{B_{1g}}}$ phonon.}
\label{fig:Raman}
\end{figure}

Raman spectra of an under- and optimally doped CLBLCO samples are shown in
Fig~\ref{fig:Raman}. The samples are freshly cleaved and cooled in an
evacuated cryostat to $T=20$~K. One surface is shown in the inset of Fig~\ref%
{fig:Raman}. The spectra are obtained by a Jobin-Yvon micro-Raman
spectrometer (LabRam-HR 800) with a $\lambda =532$ nm solid state laser and
a power of less than 1~mW to avoid overheating. The laser is focused by a
microscope objective lens with $\times $~50 magnification to a spot of about
10~$\mathrm{\mu m}$ diameter. The scattered light is collected in
backscattering geometry and dispersed by a diffraction grating with 1800
grooves/mm onto a liquid nitrogen cooled CCD array. Polarized Raman spectra
are obtained with a Polaroid polarizer and analyzer placed after the laser
and before the grating.

Since CLBLCO belongs to the tetragonal space group $\mathrm{{P_{4/mmm}}}$,
symmetry analysis \cite{bilbao} yields $4\mathrm{{A_{1g}+1B_{1g}}=5}$ Raman
active optical phonons for scattering in the $ab$ plane. In $a^{\prime
}(c,c)a^{\prime }$ polarization (i.e. both polarizer and analyzer rotated by
45$^{\circ }$ from the crystallographic $a$ axis), only $\mathrm{{A_{1g}}}$
phonons are allowed, while in $a^{\prime }(c,c)b^{\prime }$ polarization
(with the analyzer set perpendicular to the polarizer and 45$^{\circ }$ from
the crystallographic $b$ axis) only $\mathrm{{B_{1g}}}$ phonons are allowed.
In the underdoped crystal, four of the five phonons are clearly observed at
145, 298, 423, 446 $\mathrm{{cm^{-1}}}$. The fifth optical phonon is either
at 209 $\mathrm{{cm^{-1}}}$ where a very shallow peak is observed, or it
overlaps with the multi-phonon peaks starting around 500 $\mathrm{{cm^{-1}}}$%
. For comparison, YBCO$_{6}$~has five phonons \cite{phononofYBCO} at about
140, 230, 335,435, 500~$cm^{-1}$.

Using the rotational anisotropy of the 298 $cm^{-1}$ $B_{1g}$ phonon with
respect to the electric field, the $a$ and $b$ direction of the crystal are
found. The center inset of Fig.~\ref{fig:Raman} shows the phonon intensity
as a function of the polarizer angle from $a$. The crystal direction are
depicted on the cleave surface in Fig.~\ref{fig:Raman}.

Another feature of the underdoped crystal, common to all cuprates, is the
broad peak centered at $\Delta E_{max}=2580(20)\mathrm{{cm^{-1}}}$. This
peak is attributed to two-magnon scattering. The Heisenberg exchange
constant $J$ is roughly given by $\Delta E_{max}/3=1240(10)$~K \cite%
{CanaliPRB92}. This estimation is in good agreement with the $\mu $SR data
given in Fig.~\ref{fig:PhaseDiagram}(a) and (b); for the most underdoped $%
x=0.1$ sample, we have $J/T_{c}^{max}=20.4(1.5)$ and $T_{c}^{max}=58$~K,
giving $J=1170(100)$~K. Achieving a better agreement between the two values
requires further investigations. Finally, the two-magnon peak is absent from
the spectrum of the optimally doped crystal due to an overdamping of the
spin excitations, which is also depicted in Fig~\ref{fig:Raman}; the spectra
are shifted in intensity for clarity.

Our data indicate that CLBLCO can be grown as large single crystals. The
crystals are large enough for neutron scattering. The crystals are
cleavable, giving high quality surfaces, which make them suitable for
surface probes. Within the realm of our experiments, the crystals have the
same properties as the original powder. These properties, combined with the
unique phase diagram of CLBLCO open new perspectives for cuprate research.

Acknowledgments: One of us (AK) would like to thank John Tranquada and Genda
Gu for a two month visit to their lab and excellent tutorial on single
crystal growth, and to Kazimierz Conder and Ekaterina Pomjakushina for
helpful discussion. This research was funded by the Israeli Science
Foundation and the \textbf{ESF}.

\section{Appendix}

The crystals are grown using a Crystal System Corp. optical furnace
(FZ-T-4000-H) equipped with four parabolic mirrors and 300W lamps. CLBLCO
powder is first prepared using solid state reaction from stoichiometric
proportions of La$_{2}$O$_{3}$ (99.99\%), CaCO$_{3}$ (99.9\%), BaCO$_{3}$
(99.9\%) and CuO(99\%). The starting materials are mixed, calcined, and
ground repeatedly at 900, 925, 950, 950~C$^{\circ }$ for 24 hours, then
tested for impurities using XRD. The CLBLCO powder is then packed into
rubber tubes and hydrostatically pressed at 400~MPa, producing 15-20~cm $%
\times \phi $~ 7~mm cylindrical rods. The density of the rods reaches more
than 90\% after sintering them at 1040~C$^{\circ }$ for 24 hours ensuring
stable uninterrupted growth. The growth is carried out at a rate of
0.35~mm/h in argon atmosphere with 0.05-0.1\% oxygen, with both feed and
seed rotating in opposite directions at 15~rpm. The growth lasts up to 10
days resulting in 50-100~mm long crystals. The grown crystals are annealed
in argon at 850~C$^{\circ }$ for 120 hours to relieve thermal stress, and,
when needed, to remove excess oxygen for magnetic measurement ($\mu $SR and
elastic neutron scattering). It is important to mention that we do not find
any traces of copper on the furnace quartz tube as usually happens with LSCO
growth. Therefore, we do not add any excess copper to the initial powder.

We also perform Inductively-coupled plasma atomic-emission spectroscopy
(ICP-AES)\cite{ICP} measurements on the crystals and the starting powders.
For ICP we prepare 10 test tubes with the 100~mg of crystal dissolved in
"Trace Select" grade Nitric Acid, as well as 5 test tubes of the starting
powder. ICP allows us to measure the molar ratio of Ca, Ba, and La to Cu in
the mixture. Knowing that we have 3 Cu atoms per unit cell, we found Ca $%
=0.102\pm 0.005$, Ba $=1.565\pm 0.05$ and La $=1.319\pm 0.02$ atoms per unit
cell in the powder case. The nominal values should be Ca $=0.1$, Ba $=1.65$
and La $=1.25$. This sets the sensitivity of the ICP technique. For the
crystal we find that Ca $=0.102\pm 0.005$, Ba $=1.57\pm 0.02$ and La $%
=1.295\pm 0.006$ atoms per unit cell. Therefore, there is no difference in
the Ca, Ba, or La content, between the crystals and powders within the
accuracy of the measurement.

\end{document}